\documentclass[]{ws-p9-75x6-50}

\def\apj{{\em The Astrophysical Journal}}

\def\be{\begin{equation}}
\def\ee{\end{equation}}
\def\bea{\begin{eqnarray}}
\def\eea{\end{eqnarray}}

\def\spose#1{\hbox to 0pt{#1\hss}}
\def\approxlt{\mathrel{\spose{\lower 3pt\hbox{$\sim$}}
        \raise 2.0pt\hbox{$<$}}}
\def\approxgt{\mathrel{\spose{\lower 3pt\hbox{$\sim$}}
        \raise 2.0pt\hbox{$>$}}}

\begin{document}

\title{GENERAL RELATIVISTIC EFFECTS ON MAGNETAR MODELS OF AXPs} 

\author{F. \"OZEL}

\address{Harvard-Smithsonian Center for Astrophysics, 60 Garden St., 
\\ Cambridge, MA 02138, USA\\E-mail: fozel@cfa.harvard.edu}   

\maketitle\abstracts{
General relativistic bending of light dramatically alters the
variability of X-ray emission originating from the surfaces of
ultramagnetic neutron stars. We construct radiative equilibrium models
of such strongly magnetic cooling neutron stars with light-element
atmospheres to compute the angle- and energy-dependent intensity
emerging from their surfaces and find that the beaming of surface
emission is predominantly non-radial. The combination of this
radiation pattern with the calculations of light bending yields pulse
amplitudes that vary non-monotonically with the neutron star
compactness and the size of the emitting region. The significant
suppression of the pulse amplitude for large emitting areas provides
very strong constraints on the mechanisms that can simultaneously
produce high periodic variability and X-ray luminosity. We apply these
results to the thermally-emitting magnetar models of anomalous X-ray
pulsars (AXPs), which are bright slowly-rotating X-ray sources with
large pulse amplitudes. We use the observed fluxes and pulse
amplitudes for all known AXPs and show that thermal emission from two
antipodal regions on their surfaces, as predicted by some magnetar
models, is inconsistent with these observed properties. }

Since their identification as a class in 1995, Anomalous X-ray Pulsars
(AXPs), distinguished primarily by their X-ray brightness, by
pulsations with 6-12 s periods and by the lack of radio or optical
counterparts [1], have been elusive sources challenging our current
theoretical understanding. The combination of their soft spectra (with
color temperatures of $\sim 0.5$~keV), their X-ray luminosities of
$10^{34-36} {\rm erg\, s}^{-1}$ and their large pulse amplitudes
ranging between $15\%-70\%$, have also been important properties to
account for within the current models of AXPs.

One such class of models, commonly referred to as the magnetar models,
relies on anisotropic thermal emission from the surface an
ultramagnetic ($B \approxgt {\rm few} \times 10^{13} {\rm G} $)
neutron star (NS) as the source of the modulated X-ray emission of
AXPs, powered either by the decay of this strong field in the crust or
by the latent heat from the young NS core. This surface emission is
thought to give rise to the soft and dominant component in the spectra
of AXPs and, therefore, primarily determines the spectral and timing
properties of these sources.

\begin{figure}[t] 
\centerline{ \psfig{figure=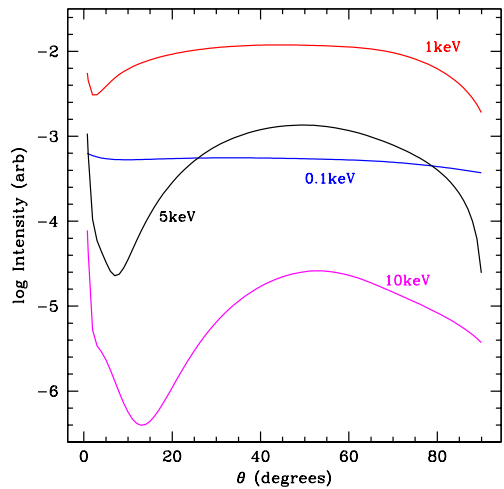,width=7.0cm} }
\caption{The beaming of radiation emerging from a NS atmosphere
with $B= 10^{14}$ G and $T_e=0.3$ keV at photon energies $0.1, 1, 5,$
and 10 keV. \label{fig:fig1}}
\end{figure}

We carried out the first angle- and energy-dependent radiative
transfer calculations at ultrastrong magnetic fields and constructed
radiative equilibrium models of cooling NSs [2]. We assume that the NS
atmosphere is a fully ionized H plasma with an ideal gas equation of
state in plane-parallel geometry, and that the magnetic field is
normal to the surface. The problem then consists of solving the angle-
and energy-dependent equation of radiative transfer coupled to the
equation of hydrostatic balance and subject to the constraint of
radiative equilibrium. We take into account bremsstrahlung and fully
angle-dependent conservative scattering processes at high magnetic
fields. We employ a modified Feautrier method for the solution of the
radiative transfer problem to allow for the two polarization modes of
the photons in the plasma, and a partial linearization scheme based on
a Uns\"old-Lucy temperature correction method to determine the
thermodynamic structure of the atmospheres in radiative
equilibrium. Figure~\ref{fig:fig1} shows the beaming of emerging
radiation at photon energies of 0.1, 1, 5, and 10 keV. At almost all
magnetic field strengths and photon energies, the beaming has a radial
(pencil) and prominent broad non-radial (fan) component. At $B
\approxgt 10^{14} {\rm G}$, the radial component is significant only
at high photon energies. The X-ray spectra of the emerging radiation
(not shown) are broader than blackbody spectra and, allowing for hard
excess, can be fit with a blackbody of color temperature $T_c \approx
1.2-1.8~T_e$, where $T_e$ is the effective temperature of the
atmosphere.

General relativistic (GR) effects dramatically alter the observable
modulation of X-ray emission originating from a NS surface due to the
curved photon paths in strong gravitational fields. For radially
peaked beaming patterns, the result is a significant suppression of
the pulse amplitude because the curved trajectories effectively allow
a larger fraction of the NS surface to be visible to the observer at
any pulse phase. Taking into account such effects and considering an
antipodal emission geometry, it was shown that the $\approxgt 50 \%$
pulse amplitudes observed in 3 AXPs limit the angular sizes of the
surface emitting areas to $\rho \approxlt 20$ degrees for strongly
radially-peaked beaming of the emerging radiation [3].

The case of the non-radial beaming that is relevant for surface
emission from a magnetar is significantly more complex, and in
particular the pulse amplitude varies non-monotonically with the
neutron star compactness and may also not be monotonic with the size
of the emitting area (Fig.~2). This is due to several reasons. First,
the peak of the non-radial beams appear broader and at an angle
further away from the surface normal due to GR effects. The emission
reaching the observer from two antipodal emitting regions can
therefore add at a phase $\phi$ away from the direction of the
magnetic field to yield the highest flux at $\phi=\pi/2$. This effect
is present for a range of curvatures of photon paths, corresponding to
a range of neutron star compactnesses, as well as for a range of spot
sizes. In addition, neutron stars of different compactness give rise
to different redshifts so that the $1-10$~keV range detected by an
observer at infinity corresponds to an increasingly higher photon
energy range on the neutron star surface for increasing
compactness. Therefore, for strongly energy-dependent beaming of
radiation as in the case of magnetars, the resulting pulse amplitude
over an observed energy range is strongly affected by the NS
compactness.

\begin{figure}[t] 
\begin{minipage}{7.0cm}
\psfig{figure=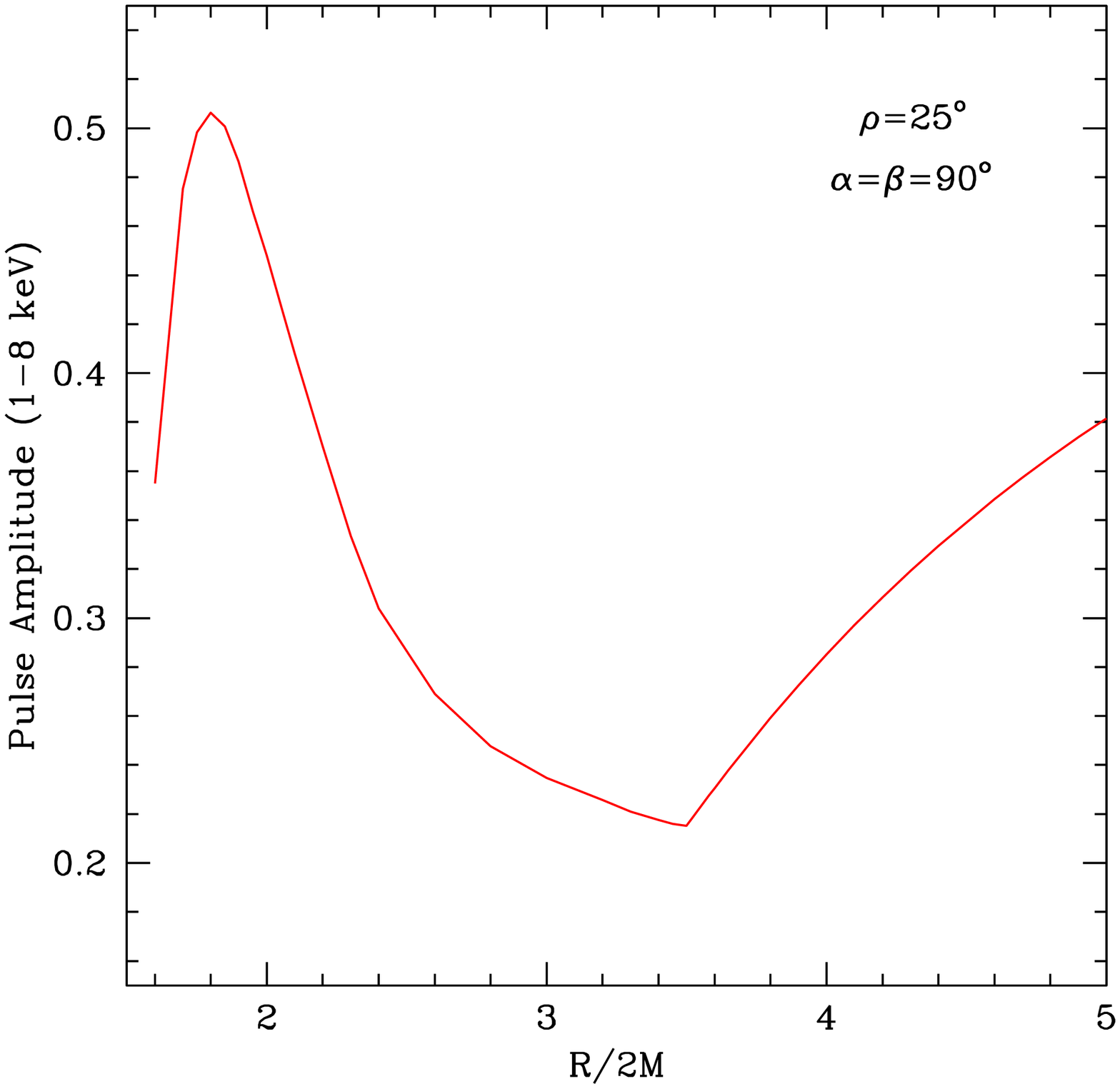,width=6.0cm}
\end{minipage}
\begin{minipage}{7.0cm}
\psfig{figure=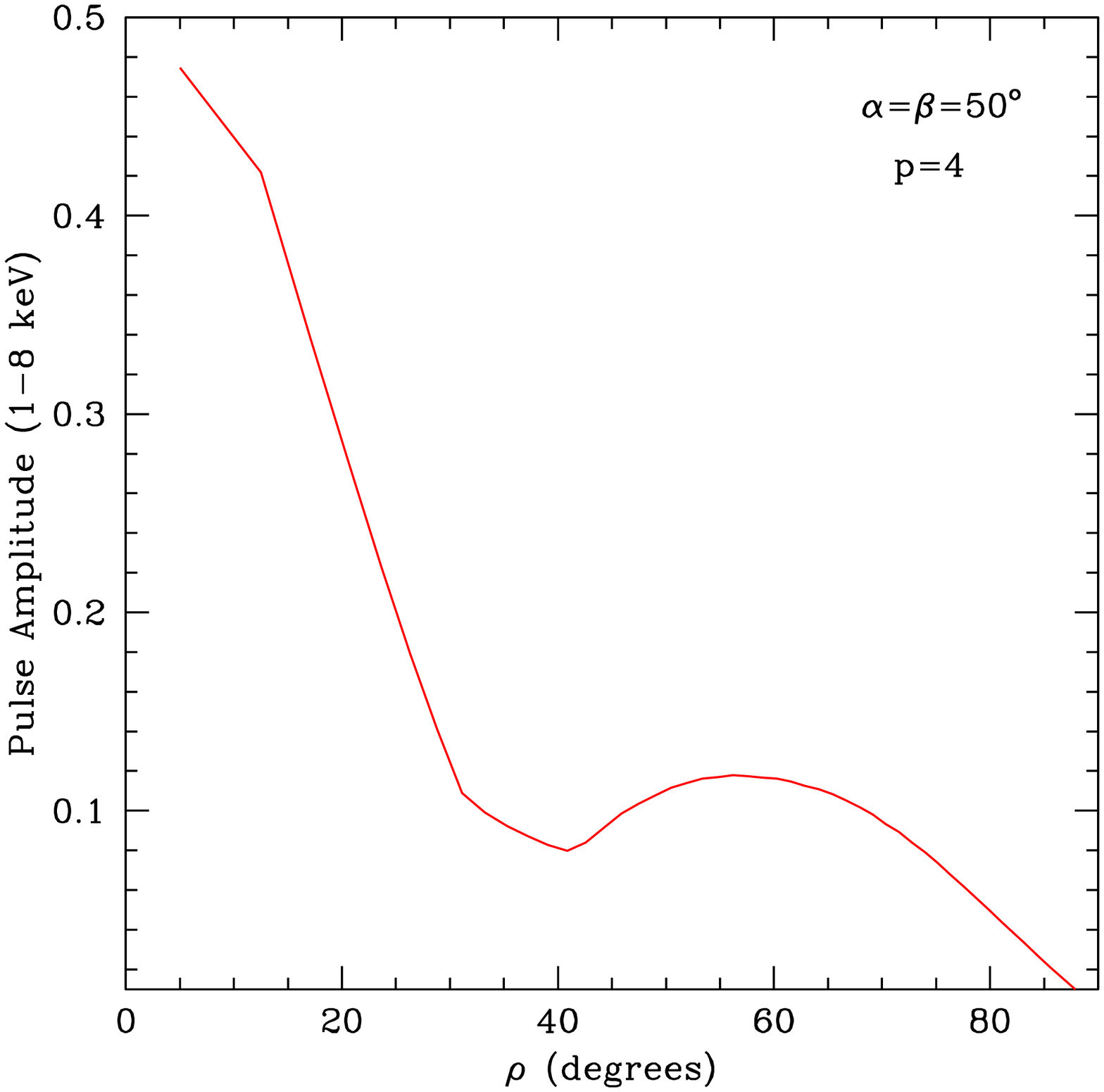,width=6.0cm}
\end{minipage}
\caption{(Left) The dependence of the pulse amplitude on the compactness 
of the neutron star, with the size of the emitting region fixed at
$\rho=25^\circ$. (Right) The dependence of the pulse amplitude on the
size of the emitting region for $R/2M=4$. In both panels, the other
model parameters are $B= 10^{14}$~G and
$T_e=0.3$~keV. \label{fig:fig2}}
\end{figure}

Figure~\ref{fig:fig2} (left panel) shows the pulse amplitude from a
$B=10^{14}$~G NS as a function of $R/2M$ which specifies the stellar
compactness.  Note that a smaller $R/2M$ represents a more compact
NS. The two angles $\alpha$ and $\beta$ which correspond,
respectively, to the position of the emitting region and of the
observer with respect to the rotation axis are fixed at $90^\circ$,
the orthogonal rotator geometry. As $R/2M$ decreases, we first obtain
the more commonly known suppression of pulse amplitudes until the
bending of photon paths is strong enough to give the maximum observed
flux at $\phi=\pi/2$. This increase in the pulse amplitude typically
starts at $R/2M \approxlt 2.8-3.8$ depending on the size of the
emitting region. For very compact NSs, nearly the entire surface
becomes visible to the observer and thus the behavior is again
reversed. Figure~\ref{fig:fig2} (right panel) shows the dependence of
the pulse amplitude on the size of the emitting region, for $R/2M =
4.0$ and $\alpha=\beta=60^\circ$. All other model parameters are as
before.  The non-monotonic variation of the pulse amplitude with
increasing size of the emitting region is due to similar reasons. 

The significant overall suppression of the pulse amplitude at large
emitting areas has important consequences for thermal emission from
the surface of a neutron star. This is because for thermally emitting
sources at a given effective temperature, a natural anticorrelation
arises between the maximum luminosity of the source and the maximum
observable pulse amplitude. This introduces a limiting curve in the
luminosity-pulse amplitude space, above which no thermally emitting
system is allowed [4].

\begin{figure}[t]
 \centerline{\psfig{file=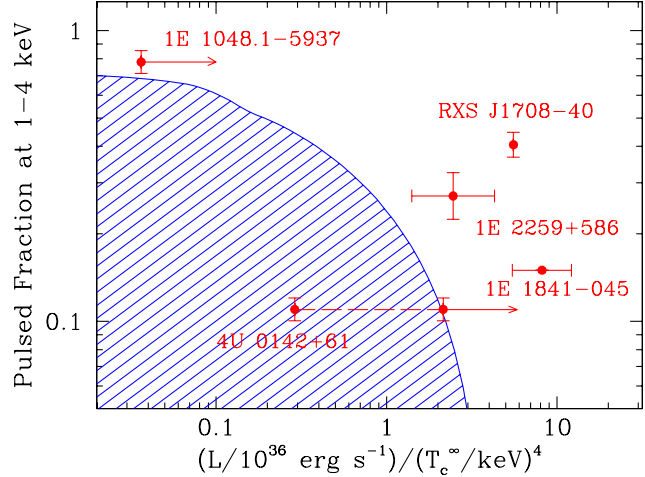,angle=-90,width=9truecm}}
\caption{The limiting curve for a thermally emitting magnetar model 
with two antipodal emitting regions in pulse amplitude--luminosity
space. Four out of the five shown AXPs lie outside the allowed
region. \label{fig:fig3}}
\end{figure}

Combining the results of the radiative transfer calculations with
general relativistic photon transport, we show in
Figure~\ref{fig:fig3} the maximal curve allowed by the thermal
emission models on the pulse amplitude-luminosity space. In obtaining
this curve, we vary all model parameters including the magnetic field
strength, effective temperature, and the NS compactness. On the same
diagram, we also show the observed fluxes and pulse amplitudes for the
five AXPs, making use of the best distance estimates to these sources
to calculate source luminosities. Four out of the five sources for
which data are available lie well outside of the region allowed by
thermal magnetar models, rendering surface emission from ultramagnetic
NS with two antipodal regions inconsistent with observations.

\begin{thebibliography}{99}

\bibitem{a} S. Mereghetti, in proceedings of 
The Neutron Star - Black Hole Connection, in press (astro-ph/9911252)

\bibitem{b} F. \"Ozel, \apj, submitted (astro-ph/0103227) 

\bibitem{c} S. DeDeo, D. Psaltis, and R. Narayan, \apj, in press 
(astro-ph/0004266)

\bibitem{d} F. \"Ozel, D. Psaltis, and V. Kaspi, \apj, submitted 
(astro-ph/0105372)

\end{thebibliography}
\end{document}